%Paper: hep-ph/9303311
%From: ULF MEISSNER <MEISSNER@ITP.unibe.ch>
%Date: Mon, 29 Mar 1993 17:02:25 CET

\magnification = 1200
\def\lapp{\hbox{$ {     \lower.40ex\hbox{$<$}
                   \atop \raise.20ex\hbox{$\sim$}
                   }     $}  }
\def\rapp{\hbox{$ {     \lower.40ex\hbox{$>$}
                   \atop \raise.20ex\hbox{$\sim$}
                   }     $}  }

\def\krig#1{\vbox{\ialign{\hfil##\hfil\crcr
           $\raise0.3pt\hbox{$\scriptstyle \circ$}$\crcr\noalign
           {\kern-0.02pt\nointerlineskip}
%          {\kern-0.06pt\nointerlineskip}
$\displaystyle{#1}$\crcr}}}
\def\upar#1{\vbox{\ialign{\hfil##\hfil\crcr
           $\raise0.3pt\hbox{$\scriptstyle \leftrightarrow$}$\crcr\noalign
           {\kern-0.02pt\nointerlineskip}
$\displaystyle{#1}$\crcr}}}
\def\ular#1{\vbox{\ialign{\hfil##\hfil\crcr
           $\raise0.3pt\hbox{$\scriptstyle \leftarrow$}$\crcr\noalign
           {\kern-0.02pt\nointerlineskip}
$\displaystyle{#1}$\crcr}}}

\def\Tr{\,{\rm Tr }\,}

\def\g5{\gamma_5}

\topskip=0.60truein
\leftskip=0.18truein
\vsize=8.8truein
\hsize=6.5truein
\tolerance 10000
\hfuzz=20pt

\baselineskip 12pt plus 1pt minus 1pt
\pageno=0
\centerline{\bf CRITICAL ANALYSIS OF BARYON MASSES AND SIGMA-TERMS}
\smallskip
\centerline{{\bf IN HEAVY BARYON CHIRAL PERTURBATION THEORY
 }\footnote{*}{Work supported in part by Deutsche Forschungsgemeinschaft
and by Schweizerischer Nationalfonds.\smallskip}}
\vskip 24pt
\centerline{V\'{e}ronique Bernard}
\vskip  4pt
\centerline{\it Centre de Recherches Nucl\'{e}aires et Universit\'{e}
Louis Pasteur de Strasbourg}
\centerline{\it Physique Th\'{e}orique, Bat. 40A,
BP 20, 67037 Strasbourg Cedex    2,
France}
\vskip 12pt
\centerline{Norbert Kaiser}
\vskip  4pt
\centerline{\it Physik Department T30,
Technische Universit\"at M\"unchen}
\centerline{\it
James
Franck Stra{\ss}e,
W-8046 Garching, Germany}
\vskip 12pt
\centerline{Ulf-G. Mei{\ss}ner\footnote{$^\dagger$}{Heisenberg
Fellow.}}
\vskip  4pt
\centerline{\it Universit\"at Bern,
Institut f\"ur Theoretische Physik}
\centerline{\it Sidlerstr. 5, CH--3012 Bern,\ \ Switzerland}
\vskip 0.5in
\centerline{\bf ABSTRACT}
\medskip
We present an analysis of the octet baryon masses and the $\pi N$ and $KN$
$\sigma$--terms in the framework of heavy baryon chiral perturbation theory.
At next-to-leading order, ${\cal O}(q^3)$, knowledge of the baryon masses and
$\sigma_{\pi N}(0)$ allows to determine the three corresponding finite
low--energy constants and to predict the the two $KN$ $\sigma$--terms
$\sigma^{(1,2)}_{KN} (0)$. We also include the spin-3/2 decuplet in the
effective theory. The presence of the non--vanishing energy scale due to the
octet--decuplet splitting shifts the average octet baryon mass by an infinite
amount and leads to infinite renormalizations of the low--energy constants.
The first observable effect of the decuplet intermediate states to the
baryon masses starts out at order $q^4$. We argue that it is not sufficient to
retain only these but no other higher order terms to achieve a consistent
description of the three--flavor scalar sector of baryon CHPT. In addition, we
critically
discuss an SU(2) result which allows to explain the large shift of
$\sigma_{\pi N}(2M_\pi^2) - \sigma_{\pi N}(0)$ via intermediate $\Delta
(1232)$ states.
\vfill
\noindent BUTP--93/05\hfill March 1993

\noindent CRN--93--06
\eject
\baselineskip 12pt plus 1pt minus 1pt
\noindent{\bf I. \quad INTRODUCTION}
\medskip
At low energies,
chiral symmetry governs the interaction of the low-lying
hadrons. To a good first approximation the current masses of the three light
quarks can be set to zero and one can expand the QCD Green functions
in powers of external momenta
{\underbar {and}} quark masses around the so-called chiral limit. Assuming that
the order parameter of chiral symmetry breaking, $B_0=-<0|\bar u u|0>/F_\pi^2$,
is of order 1 GeV, an unambiguous scheme emerges in the meson sector [1].  In
the baryon sector matters are more complicated. The low-lying baryons have a
non-vanishing mass in the chiral limit,
of the order of the chiral symmetry
breaking scale $B_0$, which complicates the low-energy structure in the
meson-baryon system considerable [2]. Making use of methods borrowed from heavy
quark effective field theories, Jenkins and Manohar [3] proposed to consider
the baryons as very heavy static sources (see also Gasser and Leutwyler [4] and
Weinberg [5]). This allows to define velocity eigen-fields and to leading
order the troublesome baryon mass term can be eliminated from the Dirac
lagrangian. This procedure is completely equivalent to the well-known
Foldy-Wouthuysen transformation used in QED to order operators by inverse
powers of the mass of the Dirac field\footnote{*}{For a systematic analysis in
flavor SU(2), see ref.[6].}. A further complication in the baryon sector is the
closeness of the first resonance multiplet. While in the meson sector the
vector mesons only appear above 770 MeV, the mass of the $\rho$-meson, the
spin-3/2 decuplet is only separated by 230 MeV (in average) from the spin-1/2
ground state octet. This raises the immediate question whether these spin-3/2
fields should be included in the effective field theory from the very
beginning. Also ample phenomenological evidence from the nucleon sector exists
which points towards the importance of the spin-3/2 fields. In ref.[7] the
effective field theory was enlarged to include the decuplet. Furthermore,
numerous calculations have been performed concerning baryon masses, hyperon
non-leptonic decays [9], nucleon polarizabilities [10] and so on
within this framework.
All these calculations are done under the assumption that only the terms
non-analytic in the quark masses
are important and therefore carry a spurious
dependence on the scale introduced in dimensional regularization. In addition,
kaon and eta loop contributions can be large, as first pointed out by
Bijnens, Sonoda and Wise [11]. However, it is argued that the decuplet
contributions to a large extent cancel these large kaon and eta loop terms,
leaving one with a fairly well behaved chiral expansion even in the three
flavor case.

While one might content oneself with these rather positive results, a closer
look at how they are obtained makes one feel uneasy about them. First, chiral
perturbation theory is a systematic expansion meaning that to a given order one
has to take into account {\underbar{all}} contributions, may they be from loops
(and thus eventually non-analytic in the quark masses) or higher order contact
terms. These are multiplied by a priori unknown
coefficients, the so-called low-energy constants. Only the sum of loop and
higher order contact terms is naturally scale independent. Second, many of the
results based on the approach of taking only calculable loop diagrams with
octet and decuplet intermediate states only give consistent results if one uses
the rather small $D$ and $F$ axial vector coupling proposed in ref.[7]. These
values, to our opinion, are simply an artefact of the calculational procedure
and do not reflect the results of a consistent chiral perturbation calculation.

To put the finger on the problem, we reconsider here the classical problem of
the baryon masses and $\sigma$-terms, namely the $\pi N$ $\sigma$-term and two
$KN$ $\sigma$-terms. This is essentially the scalar sector of baryon chiral
perturbation theory. The first systematic analysis was performed by Gasser [12]
and the long-standing problem with the $\pi N$ $ \sigma$-term has recently been
resolved by Gasser, Leutwyler and Sainio [13]. They combined a
dispersion-theoretical approach with chiral symmetry constraints to show that
the $\sigma$-term shift $\sigma_{\pi N} (2M_\pi^2) - \sigma_{\pi N} (0) $ is as
large as 15 MeV. Therefore the empirical value $\sigma_{\pi N} (2M_\pi^2)
\simeq $  60 MeV determined
from $\pi N$ scattering can be reconciled if the strange
quark admixture in the proton is $y \simeq  0.2$, leading to a mass shift of
$m_s <p|\bar s s|p> \simeq 130$ MeV, considerably lower than expected from
first order quark mass perturbation theory [12,14] and anticipated in some
models of dynamical chiral symmetry breaking [15]. Jenkins and Manohar have
argued that the heavy fermion formulation with decuplet fields is indeed
consistent with these values [16]. This, in fact, is the statement we wish to
elaborate on. First we perform a complete calculation up to order $q^3$, which
only involves intermediate octet states. At this order (one-loop approximation)
one has three counterterms with a priori unknown but finite coefficients. These
can be fixed from the octet masses $(m_N, m_\Lambda, m_\Sigma, m_\Xi)$ and the
value $\sigma_{\pi N}(0) $ since one of the counter terms appears in the baryon
mass formulae in such a way that it always can be lumped together with the
average octet mass in the chiral limit. This allows us to predict
the two $KN$ $\sigma$-terms, $\sigma_{KN}^{(1)}(0)$ and $\sigma^{(2)}_{KN}(0)$
as well as
the $\sigma$-term shifts to the respective Cheng-Dashen points and the
matrix element $m_s <p|\bar ss |p>$.   We then proceed and add the low-lying
decuplet fields. Here we leave the consistent calculation since the first
visible effect of the decuplet appears at order $q^4$. So, in principle we
should account for a host of other terms. Our aim is, however, more modest. We
simply want to check whether the assertions made in refs.[7-10,16] can be
considered    sound. The first, and rather obvious, observation to be made is
that the inclusion of the decuplet fields spoils the consistent chiral power
counting. This
can be traced back to the residual mass dependence which can not be removed
from the lagrangian involving velocity-dependent fields. Stated differently, in
the chiral limit the average decuplet and octet masses differ by a nonzero
amount of a few hundred MeV. This leads to complications similar to the ones in
the relativistic formulation of baryon chiral perturbation theory (related to
the finite value of the nucleon mass in the chiral limit) discussed in ref.[2].
Also, in the baryon mass spectrum one finds an infinite renormalization of the
previously finite counter terms of chiral power two. These problems have not
yet been spelled out in the literature. For the numerical results we refer the
reader to later sections, where it will become clear that one has to work
harder to get a consistent picture of heavy baryon CHPT beyond $q^3$. As a nice
by-product, we find an SU(2) result which allows to explain the large shift
$\sigma_{\pi N}(2M^2_\pi) - \sigma_{\pi N} (0)$ in terms of the $\Delta(1232)$
states (modulo the many other unknown effects appearing at order $q^4$ or
higher). It is important to differentiate between the pure SU(2) results (where
the kaons and etas only contribute very little) and the SU(3) results, which
are afflicted by potentially large cancellations between $K$ and $\eta$ loops
on one side and decuplet contributions on the other side. It is conceivable
that one should first address these genuine SU(2) results and understand the
chiral expansion for them before one turns to the more complicated three flavor
sector. A first step was done in ref.[6] and we will come back to this problem
in a future publication. To close the introduction, let us point out that a
review on baryon CHPT is available [17], another one on the heavy fermion
formulation [18] and related aspects are discussed in the more recent
CHPT review [19].
\bigskip
\noindent{\bf II. \quad FORMALISM}
\medskip
\noindent{\bf II.1. Baryon masses and
$\sigma$--terms at next--to--leading order}
\medskip
In this section, we will give the formalism necessary to discuss the baryon
masses and $\sigma$-terms. Our starting point is the effective chiral
lagrangian of the pseudoscalar Goldstone bosons coupled to octet baryons (we do
not exhibit the standard meson lagrangian)
$${\cal L}^{(1)}_{\phi B}= \Tr(\bar B \,i v\cdot{\cal D} \,B) + D\, \Tr (\bar B
S^\mu \{ u_\mu , B\} ) + F \, \Tr ( \bar B S^\mu [ u_\mu, B] ) \eqno(2.1)$$
where the Goldstone fields $\phi $ are collected in the $SU(3)$ matrix
$$U(x) = \exp[i \phi(x)/F_p], \qquad u(x) = \sqrt{U(x)}, \qquad u_\mu
= i u^\dagger \nabla_\mu U u^\dagger \eqno(2.2)$$
and $F_p$ is the pseudoscalar decay constant in the chiral limit. $B$ is the
standard $SU(3)$ matrix representation of the low-lying spin-1/2 baryons
($p,n,\Lambda,\Sigma^0 , \Sigma^\pm, \Xi^0, \Xi^-$) and $S_\mu $ is a covariant
spin operator satisfying $v\cdot S =0$ and $S^2 = (1-d) /4$ in $d$ space-time
dimensions.  The two axial coupling constants $D$ and
$F$ are subject to the constraint $D+F = g_A = 1.26$. We work in the heavy mass
formalism, which means that baryons are considered as static sources and
equivalently their momenta decompose as
$$p_\mu = m_0 \, v_\mu + l_\mu \eqno(2.3)$$
with $m_0$ the average octet mass in the chiral limit, $v_\mu$ the baryon
four-velocity ($v^2=1$) and $l_\mu$ a small off-shell momentum. In this extreme
non-relativistic limit one can define velocity dependent fields such that the
troublesome baryon mass term disappears from the original Dirac lagrangian for
the baryons. In the absence of the baryon mass term a consistent low-energy
expansion can be derived. The baryon propagator reads $i/(v\cdot l +
i\epsilon)$. The low energy expansion resulting from loops goes along with an
expansion in inverse powers of the baryon mass $m_0$. As indicated by the
superscript (1) in eq.(2.1), tree level diagrams calculated from the lowest
order effective lagrangian are of order $q$, with $q$ denoting a genuine small
momentum. At next-to-leading order one has to include the one-loop graphs using
solely the vertices given by ${\cal L}^{(1)}_{\phi B}$ and additional chirally
symmetric counterterms of order $q^2$ and $q^3$, since the one-loop graphs all
have chiral power $q^3$. These contact terms are accompanied by a priori
unknown coupling constants and have to be fixed phenomenologically. In the
present context only counter terms of chiral power $q^2$ contribute which
account for quark mass insertions,
$${\cal L}_{\phi B}^{(2)} = b_D\, \Tr (\bar B \{ \chi_+ , B \} ) + b_F \, \Tr
(\bar B [ \chi_+,B]) + b_0 \, \Tr (\bar BB ) \Tr (\chi_+) \eqno(2.4)$$
with $\chi_+ = u^\dagger \chi u^\dagger + u \chi^\dagger u$ and $\chi = 2 B_0 (
{\cal M} + {\cal S})$ where $\cal S$ denotes the nonet of external scalar
sources. As we will see later on, the constants $b_D$, $b_F$ and $b_0$ can be
fixed from the knowledge of the baryon masses and the $\pi N$ $\sigma$-term (or
one of the $KN$ $\sigma$-terms). The constant $b_0$ can not be determined from
the baryon mass spectrum alone since it contributes to all octet members in the
same way. To this order in the chiral expansion, any baryon mass takes the
form
$$m_B = m_0 - {1\over 24 \pi F_p^2}\bigl[ \alpha^\pi_B M_\pi^3 + \alpha^K_B
M_K^3 + \alpha^\eta_B M_\eta^3 \bigr] + \gamma^D_B b_D + \gamma^F_B b_F
-2b_0(M_\pi^2 + 2M_K^2) \eqno(2.5)$$
The first term on the right hand side
of eq.(2.5) is the average octet mass in the chiral
limit, the second one
comprises the Goldstone boson loop contributions and the third
term stems from the counter terms eq.(2.4). Notice that the loop
contribution is
ultraviolet finite and non-analytic in the quark masses since $M_\phi^3 \sim
m_q^{3/2}$. The constants $b_D$, $b_F$ and $b_s$ are therefore finite. The
numerical factors read
$$\eqalign{
\alpha_N^\pi & = {9\over 4}(D+F)^2, \quad
\alpha_N^K = {1\over 2}(5D^2 - 6DF +9F^2), \quad
\alpha_N^\eta = {1\over 4}(D-3F)^2; \cr
\alpha_\Sigma^\pi &  = D^2+6F^2, \quad
\alpha_\Sigma^K = 3(D^2+F^2), \quad
\alpha_\Sigma^\eta = D^2; \cr
\alpha_\Lambda^\pi &  = 3D^2, \quad
\alpha_\Lambda^K  = D^2  +9F^2, \quad
\alpha_\Lambda^\eta   = D^2; \cr
\alpha_\Xi^\pi & = {9\over 4}(D-F)^2, \quad
\alpha_\Xi^K   = {1\over 2}(5D^2 + 6DF +9F^2), \quad
\alpha_\Xi^\eta = {1\over 4}(D+3F)^2; \cr
\gamma_N^D & = -4M_K^2,\quad
\gamma_N^F    =  4M_K^2-4M^2_\pi; \quad \quad
\gamma_\Sigma^D   =  -4M_\pi^2, \quad
\gamma_\Sigma^F  =  0; \cr
\gamma_\Lambda^D & =  -{16\over 3}M_K^2 + {4\over3}M_\pi^2, \quad
\gamma_\Lambda^F = 0; \quad \quad
\gamma_\Xi^D =  -4M_K^2, \quad \gamma_\Xi^F =  -4M_K^2+4M^2_\pi. \cr}
\eqno(2.6)$$
At this order, the deviation from the Gell-Mann-Okubo formula reads
$$\eqalign{{1\over4} \bigl[ 3m_\Lambda + m_\Sigma - 2 m_N - 2 m_\Xi\bigr]  & =
{3F^2 - D^2 \over 96 \pi F_p^2}\bigl[ M_\pi^3 - 4 M_K^3 + 3 M_\eta^3 \bigr] \cr
& = {3F^2 - D^2\over 96\pi F_p^2} \bigl[ M_\pi^3 - 4 M_K^3 + {1\over \sqrt 3}
(M_K^2 - M_\pi^2)^{3/2}\bigr] \cr} \eqno(2.7)$$
where in the second line we have used the GMO relation for the $\eta$-meson
mass, which is legitimate if one works at next-to-leading order.

Further information on the scalar sector is given by the scalar form factors or
$\sigma$-terms which measure the strength of the various matrix-elements $m_q\,
\bar q q$ in the proton. One defines:
$$\eqalign{
\sigma_{\pi N}(t) &= \hat m <p'|\bar u u + \bar d d| p>\cr
\sigma_{KN}^{(1)}(t)&={1\over 2}(\hat m + m_s) <p'|\bar u u + \bar s s | p> \cr
\sigma^{(2)}_{KN}(t) &=  {1\over 2}(\hat m + m_s) <p'|-\bar u u + 2 \bar d d +
\bar s s| p> \cr } \eqno(2.8)$$
with $t = (p'-p)^2$ the invariant momentum transfer squared
and $\hat m = (m_u +
m_d)/2$ the average light quark mass. At zero momentum transfer, the strange
quark contribution to the nucleon mass is given by
$$m_s <p|\bar s s |p> = \biggl({1\over 2} - {M_\pi^2 \over 4M_K^2} \biggr)
\biggl[3\sigma^{(1)}_{KN}(0)+ \sigma_{KN}^{(2)}(0) \biggr]+ \biggl({1\over 2} -
{M_K^2 \over M_\pi^2 } \biggr) \sigma_{\pi N}(0) \eqno(2.9)$$
making use of the leading order meson mass formulae $M_\pi^2 = 2 \hat m B_0 $
and $M_K^2 = (\hat m + m_s) B_0$ which are sufficiently accurate to the order
we are working. The chiral expansion at next-to-leading order for the
$\sigma$-terms reads
$$ \sigma_{\pi N}(0) = {M_\pi^2 \over 64 \pi F_p^2}
\biggl[
- 4 \alpha_N^\pi M_\pi
- 2 \alpha_N^K M_K - {4\over3} \alpha_N^\eta M_\eta \biggr]
- 2M_\pi^2 (b_D +
  b_F +   2 b_0)  \eqno(2.10a)$$
$$\eqalign{ \sigma^{(j)}_{K N}(0)= & {M_K^2 \over 64 \pi F_p^2}\biggl[- 2
\alpha_N^\pi
M_\pi -3\xi^{(j)}_K M_K - {10\over3} \alpha_N^\eta M_\eta \cr &\quad - 2
\xi^{(j)}_{\pi \eta}   \alpha_N^{\pi \eta}   {M_\pi^2 + M_\pi M_\eta +
M_\eta^2 \over M_\pi + M_\eta}
\biggr]
+ 4M_K^2 (\xi^{(j)}_D\, b_D + \xi^{(j)}_F\,
b_F - b_0) \cr} \eqno(2.10b)$$
 for $j = 1,2$ with coefficients
$$\eqalign{\xi^{(1)}_K & = {7\over 3} D^2 - 2DF + 5F^2, \quad \xi^{(2)}_K =
3(D-F)^2, \quad \xi^{(1)}_{\pi \eta} = 1, \quad \xi^{(2)}_{\pi\eta} = -3, \cr
\xi^{(1)}_D  & = -1, \quad \xi^{(2)}_D = 0, \quad \xi^{(1)}_F = 0, \quad
\xi^{(2)}_F =  1; \quad \alpha_N^{\pi \eta} = {1 \over 3} (D+F)(3F-D).
\cr }\eqno(2.10c)$$
This completely determines the scalar sector at next-to-leading order. Note
that the $\pi N$ $\sigma$-term is given as $\sigma_{\pi N}(0) = \hat m\,
(\partial m_N/ \partial \hat m)$ according to the Feynman-Hellman theorem. The
shifts of the $\sigma$-terms from $t=0$ to the respective
Cheng-Dashen points  do  not
involve any contact terms,
$$\eqalign{ \sigma_{\pi N}(2M_\pi^2) - \sigma_{\pi N}(0) = & { M_\pi^2 \over 64
\pi F_p^2}\biggl\{ {4 \over 3} \alpha_N^\pi
\, M_\pi \cr  & +  {2\over 3} \alpha_N^K
\biggl[ {M_\pi^2 - M_K^2 \over \sqrt 2 M_\pi } \ln {\sqrt 2 M_K + M_\pi \over
\sqrt 2 M_K - M_\pi} + M_K \biggr] \cr & + {4\over 9} \alpha_N^\eta
\biggl[ {M_\pi^2
- M_\eta^2 \over \sqrt 2 M_\pi} \ln {\sqrt 2 M_\eta - M_\pi \over \sqrt 2
  M_\eta - M_\pi } + M_\eta \biggr] \biggr\} \cr }
 \eqno(2.11a)$$
$$\eqalign{  \sigma&_{KN}^{(j)}(2M_K^2)  - \sigma_{KN}^{(j)}(0) = {M_K^2 \over
128 \pi F_p^2}  \biggl\{ {4 \over 3} \alpha_N^\pi
\biggl[
{M_K^2 - M_\pi^2 \over \sqrt 2 M_K} \biggl( \ln  {M_K +\sqrt 2M_\pi \over
M_K - \sqrt 2 M_\pi} + i \pi \biggr) + M_\pi \biggr] \cr
&+ {20\over 9} \alpha_N^\eta
         \biggl[ {M_K^2 - M_\eta^2 \over \sqrt 2 M_K}\ln
{\sqrt 2 M_\eta +  M_K \over \sqrt 2 M_\eta - M_K}   + M_\eta \biggr]  +
2\xi^{(j)}_K M_K\cr &+ \xi^{(j)}_{\pi \eta}
 \alpha_N^{\pi \eta}   \biggl[{2M_K^2 - M_\pi^2 - M_\eta^2 \over \sqrt 2 M_K}
 \ln  {\sqrt 2M_K +  M_\pi + M_\eta \over
M_\pi +M_\eta - \sqrt 2 M_K}
+ 2{M_\pi^2 + M_\eta^2
\over M_\pi + M_\eta}  \biggr] \biggr\} \cr} \eqno(2.11b)$$
Notice that the shifts of the two $KN$ $\sigma$-terms acquire an imaginary part
since the pion loop has a branch cut starting at $t
= 4M_\pi^2 $ which is below the kaon
Cheng -Dashen
point $t = 2M_K^2$\footnote{*}{Since we choose the GMO value for the
$\eta$ mass, $M_\pi + M_\eta > {\sqrt 2} M_K$, the $\pi \eta$ loop does
not contribute to the imaginary part in eq.(2.11b). For the physical
value of the $\eta$ mass this contribution is tiny compared to the pion
loop.}.
In the limit of large kaon and eta mass the result
eq.(2.11a) agrees, evidently, with the ancient
calculation of Pagels and Pardee [20]
once one accounts for the numerical error of a
factor 2 in that paper. Clearly, the $\sigma$-term shifts are non-analytic in
the quark masses since they scale with the third power of the pseudoscalar
meson masses. Our strategy will be the following: We use the empirically known
baryon masses  and the recently determined value of $\sigma_{\pi N}(0)$ [13] to
fix the unknown parameters $m_0, b_D,b_F$ and $b_0$. This allows us to predict
the two $KN$ $\sigma$-terms $\sigma_{KN}^{(j)}(0)$. The shifts of the
$\sigma$-terms are independent of this fit. Before presenting results, let us
discuss the inclusion of the low-lying spin-3/2 decuplet in the effective
theory.
\bigskip
\noindent{\bf II.2.
Inclusion of the decuplet-fields}
\medskip
The low-lying decuplet is only separated by $\Delta = 231$ MeV from the octet
baryons, which is just ${5\over 3} M_\pi$ and considerably smaller than the
kaon or eta mass. One therefore expects the excitations of these resonances to
play an important role even at low energies. This is also backed by
phenomenological models of the nucleon in which the $\Delta(1232)$ excitations
play an important role. In the meson sector, the first resonances are the
vector mesons $\rho$ and $\omega$ at about $800$ MeV, $i.e.$ they are
considerably heavier than the Goldstone bosons. It was therefore argued by
Jenkins and Manohar [7]
to include the spin-3/2 decuplet in the effective theory
from the start. Denote by $T^\mu$ a Rarita-Schwinger fields in the heavy mass
formulation satisfying
$v\cdot T =0$. The effective lagrangian of the spin-3/2
fields at lowest order reads
$${\cal L}_{\phi BT} = -i \bar T^{\mu} \, v\cdot {\cal D} \,T_\mu
+ \Delta  \, \bar T^{\mu } T_\mu + {C\over 2 } ( \bar T^{\mu }u_\mu  B + \bar
B u_\mu T^{\mu} )\,. \eqno(2.12)$$
where we have suppressed the flavor $SU(3)$ indices. Notice that there is a
remaining mass dependence which comes from the average decuplet-octet splitting
$\Delta$ which does not vanish in the chiral limit. The constant $C$ is fixed
from the decay $\Delta \to N\pi$ or the average of some strong decuplet decays.
The decuplet propagator carries the information about the mass splitting
$\Delta$ and reads
$${iP_{\mu \nu} \over v\cdot l - \Delta + i \epsilon}\eqno(2.13)$$
with
$$P_{\mu\nu} = v_\mu v_\nu - g_{\mu\nu} - 4{d-3\over d-1} S_\mu S_\nu
\eqno(2.13a)$$
in $d$ dimensions. The projector obviously satisfies the constraints $v^\mu
P_{\mu\nu} = P_{\mu \nu} v^\nu  = 0 $ and $P_\mu^\mu = -2$.
The appearance of the mass splitting $\Delta$ spoils the exact one-to one
correspondence between the loop and low-energy expansion. Two scales $F_p$ and
$\Delta$ which are both non-vanishing in the chiral limit enter the loop
calculations and they can combine in the form $(\Delta/F_p)^2$. The breakdown
of the consistent chiral counting in the presence of the decuplet is seen in
the loop contribution to the baryon mass. The loop diagrams with intermediate
decuplets states which naively count as order $q^4$ renormalize the average
octet baryon mass even in the chiral limit by an infinite amount. Therefore one
has to add a counter term of chiral power $q^0$ to keep the value $m_0$ fixed
$$\eqalign{\delta {\cal L}^{(0)}_{\phi B} & = - \delta m_0 \, \Tr (\bar B B)\cr
\delta m_0 & = {10\over 3} {C^2 \Delta^3 \over F_p^2} \biggl[ L + {1\over
16\pi^2} \bigl( \ln {2\Delta \over \lambda} - {5\over 6} \bigr) \biggr] \cr
L & = {\lambda^{d-4} \over 16 \pi^2} \biggl[ {1\over d-4} + {1\over 2} (
\gamma_E - \ln 4\pi - 1) \biggr] \cr } \eqno(2.14)$$
with $\lambda$ the scale introduced in dimensional regularization and $\gamma_E
= 0.577215...$ the Euler-Mascheroni constant. This mass shift is similar to the
one in the relativistic version of pion-nucleon CHPT, where the non-vanishing
nucleon mass in the chiral limit leads to similar complications.

Let us now turn to the baryon masses. The inclusion  of the decuplet fields has
two effects on the mass formulae eq.(2.5). First there is an infinite loop
contribution with decuplet intermediate states and, second,
an infinite renormalization
of the order $q^2$ of the
low-energy constants $b_D, b_F$ and $b_0$. Indeed this
divergent mass shift due to decuplet loops has not been treated consistently
before. To account for it, we give the following renormalization prescription
for $b_D,b_F$ and $b_0$
$$\eqalign{
b_D & = b_D^r(\lambda) - {\Delta C^2 \over 2F_p^2} L \cr
b_F & = b_F^r(\lambda) + {5\Delta C^2 \over 12F_p^2} L \cr
b_0 & = b_0^r(\lambda) + {7\Delta C^2 \over 6F_p^2} L \cr} \eqno(2.15)$$
where the finite pieces $b_{D,F,0}^r(\lambda)$ will be determined by our
fitting procedure (see below). Therefore the decuplet contributions to the
octet masses can be written in the form
$$\delta m_B = {C^2 \over 24 \pi^2 F_p^2} \biggl[ \beta^\pi_B H(M_\pi) +
\beta^K_B H(M_K) + \beta^\eta_B H(M_\eta)\biggr] \eqno(2.16)$$
with coefficients
$$\eqalign{ \quad
\beta^\pi_N & = 4, \quad \beta^K_N = 1, \quad \beta^\eta_N = 0;
\quad \beta^\pi_\Sigma = {2\over 3}, \quad \beta^K_\Sigma = {10\over 3}, \quad
\beta^\eta_\Sigma = 1; \cr \beta^\pi_\Lambda & = 3, \quad \beta^K_\Lambda = 2,
\quad \beta^\eta_\Lambda = 0; \quad \beta^\pi_\Xi = 1, \quad \beta^K_\Xi = 3,
\quad  \beta^\eta_\Xi = 1. \cr } \eqno(2.16a)$$
and
$$\eqalign{
H(M_\phi) & = \Delta^3 \ln {2\Delta \over M_\phi} + \Delta M_\phi^2 \biggl(
{3\over 2} \ln  {M_\phi \over \lambda} - 1 \biggr) - (\Delta^2 -
M_\phi^2 )^{3/2} \ln \biggl[ {\Delta \over M_\phi} + \sqrt{ {\Delta^2 \over
M_\phi^2} - 1} \biggr] \,;\,\,\, M_\phi < \Delta \cr
H(M_\phi) & = \Delta^3 \ln {2\Delta \over M_\phi} + \Delta M_\phi^2 \biggl(
{3\over 2} \ln  {M_\phi \over \lambda}  - 1 \biggr) - (M_\phi^2 -
\Delta^2)^{3/2} \arccos{\Delta \over M_\phi} \,; \quad M_\phi > \Delta\,. \cr}
\eqno(2.16b)$$
It is instructive to expand $H(M_\phi) $ for small $M_\phi$
$$H(M_\phi) = {3\over 4} \Delta M_\phi^2 \biggl(2 \ln {2\Delta \over \lambda} -
1\biggr) + {3 M_\phi^4 \over 32 \Delta} \biggl( 4 \ln {M_\phi\over 2 \Delta} -
3\biggr) +  \dots \eqno(2.17)$$
This shows that the leading contribution of the diagrams with intermediate
decuplet states is of order $M_\phi^2$, which means linear and therefore
analytic in the quark masses. This again demonstrates the problems with the
chiral power counting in the presence of a second non-vanishing scale $\Delta$.
However, this contribution has no physical effect since it can be absorbed in
the renormalized values of $b^r_{D,F,0}(\lambda)$. So the first non-trivial
effect of the decuplet states on the baryon masses appears at order $q^4$,
which means beyond next-to-leading order. This is in agreement with the
decoupling theorem [21].
Clearly, at this order there are
many other contributions. We will come back to this point later on. The
decuplet contribution to the GMO deviation reads
$${1\over 4} \bigr[3\delta m_\Lambda + \delta m_\Sigma - 2 \delta m_N - 2
\delta m_\Xi \bigr] = {C^2 \over 288\pi^2 F_p^2} \biggl[-H(M_\pi) + 4 H(M_K)- 3
H(M_\eta)\biggr] \eqno(2.18)$$
Notice that despite the appearence of the renormalization scale $\lambda$ in
the various $H(M_\phi)$, the right hand side of eq.(2.18) is indeed scale
independent due to the GMO relation for the $\eta$-mass.

Similarly, the decuplet contributes to the $\sigma$-terms at $t=0$
are given by
$$\eqalign{\delta \sigma_{\pi N}(0) & = {M_\pi^2 C^2 \over 64 \pi^2 F_p^2}
\biggl[ 8 \tilde H(M_\pi) + \tilde H(M_K) \biggr] \cr \delta
\sigma_{KN}^{(1)}(0) & = {M_K^2 C^2 \over 64 \pi^2 F_p^2} \biggl[ 4\tilde
H(M_\pi) + {4\over 3}\tilde H(M_K) \biggr] \cr
\delta \sigma_{KN}^{(2)}(0) & = {M_K^2 C^2 \over 64 \pi^2 F_p^2} \biggl[ 4
\tilde H(M_\pi) + 2\tilde H(M_K) \biggr] \cr} \eqno(2.19)$$
with
$$\eqalign{
\tilde H(M_\phi) & =  \Delta \biggl( 2\ln{M_\phi \over \lambda} - 1 \biggr) +2
\sqrt{\Delta^2 - M_\phi^2 } \ln\biggl[ {\Delta \over M_\phi} + \sqrt{ {\Delta^2
\over M_\phi^2} - 1} \biggr] \,; \qquad M_\phi < \Delta \cr
\tilde H(M_\phi) & = \Delta\biggl( 2 \ln {M_\phi \over \lambda} - 1 \biggr) -
2\sqrt{M_\phi^2 -\Delta^2 } \arccos{\Delta \over M_\phi} \,; \qquad M_\phi >
 \Delta\,.  \cr} \eqno(2.19a)$$
The contribution of the decuplet to the $\sigma$-term shifts can be most
economically represented as a dispersion integral, the appropriate imaginary
parts are collected in appendix A. However, for the later discussion, let us
consider $\sigma_{\pi N}(2M_\pi^2) - \sigma_{\pi N}(0) $ in $SU(2)$, $i.e.$
retaining
only $N$ and $\Delta(1232)$ intermediate states. One finds in this case
$$\eqalign{\sigma_{\pi N}(2M_\pi^2) - \sigma_{\pi N}(0) =  {3g_A^2
M_\pi^4\over 16 \pi^2 F_\pi^2} \int_{4M_\pi^2}^\infty & {dt \over t^{3/2
}(t-2M_\pi^2) } \biggl[ {\pi\over 4}(t-2M_\pi^2) -\Delta \sqrt{t-4M_\pi^2} \cr
 & +(t-2M_\pi^2 + 2\Delta^2)\arctan{\sqrt{t-4M_\pi^2} \over 2\Delta} \biggr]
\cr }\eqno(2.20)$$
where we used $C= {3\over 2} g_A$, coming from
the $SU(4)$ relation between the $\pi
NN $ and $\pi N \Delta$ coupling constants. This completes the necessary
formalism. It is obvious from the discussion so far that the inclusion of the
decuplet in baryon CHPT is an incomplete attempt since there are many other
terms of order $q^4$ and higher. For example, Jenkins and Manohar [8,16]
have
included tadpole diagrams with new vertices from ${\cal L}^{(2)}_{\phi B}$.
These are of order $q^4$ and can give rise to non-analytic pieces like $m_q^2
\ln m_q$. In what follows, we will not consider such diagrams but rather assume
that anything at order $q^4$ is modelled by the inclusion of the low-lying
spin-3/2 baryons. This can, of course, not substitute a full scale $q^4$
calculation including all terms at this order, but allows us to
critically examine
the role of the decuplet fields since their contribution is unique. We are
motivated by the many papers making use of the Jenkins-Manohar proposal and
want to see to what extent such an approximation is a good thing to do.
\bigskip
\vfill   \eject
\noindent{\bf III. RESULTS}
 \medskip
In this section, we will first present results for the complete $q^3$
calculation outlined in section II and then proceed to add the decuplet.
\medskip
\noindent{\bf III.1.
Results at order ${\cal O}(q^3)$}
\medskip
First, we must fix parameters. Throughout, we us $M_\pi = 138$ MeV, $M_K = 495$
MeV and $M_\eta^2 = (4M_K^2 - M_\pi^2)/3  = (566$ MeV)$^2$ as given by the GMO
relation for the pseudoscalar mesons. This is a consistent procedure since the
differences to the physical $\eta$ mass only shows up at higher order. For
the pseudoscalar decay constant, we can either use $F_\pi = 92.6 $ MeV or $F_K
= 112$ MeV. Mostly, we use an average value $F_p = (F_\pi + F_K)/2\simeq 100$
MeV. Since in all terms $F_p^2$ appears, we will vary $F_p$ from $F_\pi$ to
$F_K$ to find out how sensitive the results are to this higher order effect
(the difference of $F_\pi$ and $F_K$ in the meson sector is of order $q^4$).
Furthermore, we use $F=0.5$ and $D=0.75$, which leads to $g_A = 1.25$. Two
other sets of $D$ and $F$ values, the one of Bourquin {\it et al.}
[22], $F= 0.477$ and $D=
0.756$, and the central value of Jaffe and Manohar [23], $F=0.47$ and $D=0.81$.
The four unknowns, which are the three low-energy constants $b_D,b_F$ and $b_0$
and the average octet mass (in the chiral limit) $m_0$ are obtained from a
least square fit to the physical baryon masses ($N,\Sigma,\Lambda,\Xi$) and the
value of $\sigma_{\pi N}(0) \simeq 45$ MeV. This allows to predict
$\sigma^{(1)}_{KN}(0)$ and $\sigma^{(2)}_{KN}(0)$ and the much discussed matrix
element $m_s<p|\bar ss |p>$, $i.e.$ the contribution of the strange quarks to
the nucleon mass. We also give the value of the GMO deviation $(3m_\Lambda +
m_\Sigma - 2m_N - 2 m_\Xi)/4$, which experimentally is 6.5 MeV.

The results of this complete ${\cal O}(q^3)$ calculation are shown in table
1. The dependence on the values of the $D$ and $F$ axial vector constants is
rather weak, only in the case of the central values of ref.[23] an accidental
cancellation occurs ($D^2 \simeq 3F^2$) which makes the GMO deviation very
small. In most of the other cases one gets roughly half of the empirical
value. However, notice that it is a very small number on the typical baryon
mass scale of 1 GeV and can therefore not expected to be predicted accurately.
The strangeness matrix element
in most cases is negative and of reasonable magnitude of
about 200 MeV. Within the accuracy of the calculation, the $KN$ $\sigma$-terms
turn out to be
$$\eqalign{\sigma^{(1)}_{KN}(0) &\simeq 200 \pm 50 \,{\rm MeV}\cr
\sigma^{(2)}_{KN}(0) &\simeq 140 \pm 40 \,{\rm MeV}\cr} \eqno(3.1)$$
which is comparable to the first order perturbation theory analysis having no
strange quarks, $\sigma^{(1)}_{KN}(0) = 205$ MeV and $\sigma^{(2)}_{KN}(0) =
63$ MeV [24]. Clearly, if one varies the value of $\sigma_{\pi N}(0)$ by $\pm
10$ MeV, the results are rather different. This shows up in a value of $b_0$
which changes from $-0.62 $ to $-0.88$ GeV$^{-1}$ which has quite a dramatic
impact on the $KN$ $\sigma$-terms and the value of $m_s<p|\bar ss |p>$. For our
analysis, however, we take the central value of $\sigma_{\pi N}(0) = 45$ MeV
[13] as given. Clearly a more accurate determination of this fundamental
quantity would be very much needed. We also have performed a calculation with
$M_\eta = 549$ MeV, the results are very close to the ones for $M_\eta$ given
by the GMO relation (for the same $D,F$ and $F_p$) with the exception of the
GMO deviation for the baryon masses.

The $\sigma$-term shifts are given by $\sigma_{\pi N}(2M_\pi^2 ) -\sigma_{\pi
N}(0) = 7.4$ MeV [6,20], which is
half of the empirical value found in ref.[13]. We will come
back to this point later on. Furthermore, one finds
$$\eqalign{ \sigma^{(1)}_{KN}(2M_K^2) - \sigma^{(1)}_{KN}(0) & = (271 + i \,
303)\,{\rm MeV}\cr  \sigma^{(2)}_{KN}(2M_K^2) - \sigma^{(2)}_{KN}(0) & = ( 21 +
i \,   303)\,{\rm MeV}\cr } \eqno(3.2)$$
whose real part can be estimated simply  Re$(\sigma_{KN}^{(1)}(2M_K^2) -
\sigma_{KN}^{(1)}(0))
\simeq [\sigma_{\pi N}(2M_\pi^2) - \sigma_{\pi
N}(0)](M_K/M_\pi)^3 = 7.4 \cdot 42.2 $ MeV = 340 MeV. The rather small
real part in $\Delta \sigma_{KN}^{(2)}$ stems from the large negative
contribution of the $\pi \eta$--loop which leads to strong cancellations.
Notice the large
imaginary parts in $\sigma^{(j)}_{KN}(2M_K^2)-\sigma^{(j)}_{KN}(0) $ due to the
two--pion cut.
\bigskip
\noindent{\bf III.2.
Results with inclusion of the decuplet}
\medskip
In the case of adding the decuplet, we will first keep the value of the
pseudoscalar decay constant $F_p = 100$ MeV fixed. For the mass splitting we
use either $\Delta = 231$ MeV or $\Delta = 293$ MeV (from the $N\Delta$
splitting) and the value of $C$ is given to be 1.8 from the strong decay
$\Delta \to N\pi$ and $C=1.5$ from an overall fit to the decuplet decays [18].
The scale of dimensional regularization is chosen at $\lambda =1$ GeV [6]. It
plays of course no role in the physical results, but it should be kept in mind
that the scale dependent values of $b_{D,F,0}(\lambda)$ are given at this
scale. For the values of $D$ and $F$, we will use our central ones ($F= 0.5,
D=0.75$) and also show results with the small values of ref.[7], $D=0.56, F =
2D/3$ together with $C= 2D=1.12$ [16].

In table 2, we show the results for the full decuplet contribution according to
eqs.(2.16) and (2.19) for the baryon masses and $\sigma$-terms, respectively.
For comparison, table 3 gives the results accounting only for the leading
term at order $q^4$ arising from the decuplet intermediate states, making use
of eq.(2.17) and the expanded form of eq.(2.19). First, one notices that for
physical values of the $F$ and $D$ constants, an inconsistent picture emerges.
While the analysis of the $\pi N$ $\sigma$-term leads one to
believe that the strange quark contributes of the order of
$15\%$ to the proton
mass, this is completely different when the decuplet is included. Comparison of
table 2 and 3 shows that the bulk of this effect comes from the terms of order
$q^4$. The conclusion drawn from this exercise is that simply taking the
decuplet fields at order $q^4$ is meaningless. It still might be possible as
argued in ref.[16] that despite a large contribution at order $M_K^4 \sim q^4$,
the chiral series might converge with the small parameter $(M_K/4\pi F_\pi)^2 =
0.16$. However, it is clear that only a complete calculation at order $q^4$
(and beyond) can give a definite answer to this question. Using the small
values of $F,D$ and $C$ [7,16], we essentially recover the results of Jenkins
and Manohar. The differences stem from the fact that we did not account for the
tadpole diagrams with one insertion of ${\cal L}^{(2)}_{\phi B}$, eq.(2.4). In
the spirit of the previous remarks, this is consistent. It is interesting to
note that the results for $\sigma^{(j)}_{KN}(0)$ are quite similar to the ones
of the full $q^3$ analysis if one uses the small values of $F,D$ and $C$.

Let us now consider the $\pi N$ $\sigma$-term shift $\sigma_{\pi N}(2M_\pi^2) -
\sigma_{\pi N}(0)$. For the preferred choice $F=0.5, D  = 0.76$ and $C= {3\over
2} g_A$ based on the coupling constant relation $g_{\pi N\Delta} = 3g_{\pi N}
/\sqrt 2$, we find
$$\sigma_{\pi N}(2M_\pi^2) - \sigma_{\pi N}(0) = 15\, {\rm MeV} \eqno(3.3)$$
which agrees nicely with the empirical result of ref.[13]. It is interesting
to discuss
this result. While the leading non-analytic piece proportional to $M_\pi^3$
gives 7.4 MeV, the same amount comes from the the analogous diagram with a
$\Delta$ intermediate state. The kaon and $\eta$ loops add a meager 1.1 MeV,
$i.e.$ they are essentially negligible. This result agrees with the
phenomenological analysis of Jameson {\it et al.}.
We should stress that the spectral
distribution Im$\sigma_{\pi N}(t)/t^2$ is much less pronounced around $\sqrt t
= 600$ MeV than in ref.[13]
but has a longer tail, so the total result remains the same. The
$\Delta$-contribution mocks up the higher loop corrections of the dispersive
analysis of Gasser $et$ $al.$ [13]. A similar phenomenon is also observed in
the calculation of the nuclear forces to order $q^4$ in heavy baryon CHPT [26].
There, the intermediate range attraction comes from the uncorrelated two-pion
exchange and some four-nucleon contact terms, whereas the phenomenological
wisdom is that correlated two-pion exchange (also with diagrams involving
intermediate $\Delta$'s) is at the origin of this effect. For a more detailed
discussion, see ref.[27]. Clearly, the result eq.(3.3) should be considered a
curious accident since the correction to the leading term and the latter are of
the same magnitude. It remains to be seen how other $q^4$ effects and higher
order corrections not yet accounted for will modify eq.(3.3).
It is, however,
important to note the essential difference to the calculation of the baryon
masses and of $\sigma^{(j)}_{KN}(0)$. In $\sigma_{\pi N}(2M_\pi^2) -
\sigma_{\pi N}(0)$, the heavy meson loops are irrelevant, $i.e.$ it is an SU(2)
statement. It therefore has a better chance to survive higher order loop
corrections since the expansion parameter is $(M_\pi/4\pi F_\pi)^2 = 0.014$. Of
course, there are also extra contact terms which will have to be evaluated.
Finally, let us notice that the $KN$ $\sigma$-terms shifts are large and that
for the small values of $F,D$ and $C$  [16] $\sigma_{\pi N}(2M_\pi^2) -
\sigma_{\pi N}(0)$ is only 6.8 MeV.
\bigskip
\noindent {\bf IV.
\quad SUMMARY AND OUTLOOK}
\bigskip
We have investigated the scalar sector of three-flavor baryon chiral
perturbation theory. The baryons were treated as very massive fields, which
allows to eliminate the troublesome mass term from the lowest order effective
meson-baryon Lagrangian. Our findings can be summarized as follows:
\medskip
\item{$\bullet$}{At next-to-leading order, i.e. order ${\cal O}(q^3)$ in the
chiral expansion, one has three finite counterterms which amount to quark mass
insertions. The respective low-energy constants are denoted $b_D$, $b_F$
and $b_0$. Their values can be determined from a least square-fit to
the baryon masses and the pion-nucleon $\sigma$-term at $t=0$. This allows to
predict the two kaon-nucleon $\sigma$-terms, $\sigma_{{\rm KN}}^{(1,2)} (0)$.
The values given in eq.(3.1) are not very different from the lowest order
analysis. The shifts to the kaon Cheng-Dashen point are complex with a large
real and large imaginary part, the latter being due to the two-pion cut. These
numbers are considerably larger than the ones estimated by Gensini [28] a
decade ago.\footnote{*}{For an update on the various extractions of the
KN  $\sigma$--terms and a discussion of these results see ref.[29].}
At this order, there is a one-to-one correspondence between the
meson loop and small momentum expansion.}
\medskip
\item{$\bullet$}{We have then proceeded and added the low-lying spin-3/2
decuplet to the effective theory. We show that the new mass scale, which is
the average octet-decuplet splitting, is non-vanishing in the chiral limit and
thus induces an infinite renormalization of the baryon self-energies. This is
analogous to the infinite mass shift in the relativistic formulation of baryon
CHPT as spelled out by Gasser et al.[2]. The consistent power counting scheme
is therefore not present any longer. Similarly, there is also an infinite
renormalization of the three low-energy constant from the ${\cal O}(q^2)$
effective meson-baryon Lagrangian. Dissecting the contributions from the
diagrams with intermediate decuplet states, one finds to leading
order self-energy
contributions which are proportional to the quark masses. However, these can
be absorded entirely in the finite values of the renormalized low-energy
constants $b^r_{D,F,0}$. The first non-trivial effect of the decuplet states
on the baryon masses appears at order ${\cal O}(q^4)$.}
\medskip
\item{$\bullet$}{The numerical evaluation of the decuplet contributions to the
baryon masses and $\sigma$-terms shows a strong dependence on the values of
$F$, $D$ and $C$, the latter one being related to the strong
decuplet-octet-meson couplings. The decuplet contributions are large and for
the physical values of $F$ and $D$, one does not have a consistent picture of
the scalar sector of baryon CHPT. We disagree with the conclusion of
refs.[8,16,18] that small values of $F$ and $D$ lead to a consistent picture
at this order. First, these values stem from an incomplete calculation and,
second, at order ${\cal O}(q^4)$ there are many other diagrams which we
(and other authors) did not take into account. The diagrams with intermediate
decuplet
states contribute to all orders in $q^2$, but dominantly at ${\cal O}(q^4)$ as
comparison of tables 2 and 3 reveals. We conclude that it is not sufficient to
include the decuplet to get an accurate machinery for baryon CHPT in the
three flavor sector.} \medskip
\item{$\bullet$}{As an interesting by-product, we have found that
intermediate
$\Delta (1232)$ states give a contribution to the $\pi N$ $\sigma$-term shift
as large as the leading order result of 7.5 MeV, so in total one has 15 MeV
in agreement with the result of ref.[13]. This is an $SU(2)$ number not
affected by large kaon and eta loop contributions. However, at order
${\cal O}(q^4)$ there are other contributions not considered here which might
invalidate this result. From this we conclude that it is mandatory to first
understand in better detail the two-flavor sector of baryon CHPT before one
can hope to have a well-controlled chiral expansion including also the strange
quark.}
\bigskip
\noindent{\bf APPENDIX: IMAGINARY PARTS OF SCALAR FORM FACTORS}
\medskip
Here, we give explicit formulae for the imaginary parts to one loop of the
three proton scalar form factors defined in eq.(2.8):
$$\eqalign{ {\rm Im}\sigma_{\pi N}(t) = &\,{M_\pi^2 \over 128  F_p^2 \sqrt
t} \biggl\{
3  (D+F)^2 (t - 2M_\pi^2 ) \theta(t-4M_\pi^2) \cr & +  ({5\over 3}
D^2-2DF + 3F^2) (t-2M_K^2) \theta(t-4M_K^2) + ({D\over 3}-F)^2(t-2M_\eta^2)
\theta(t-4M_\eta^2) \cr & + {16 C^2 \over 3\pi } \biggl[ -
\Delta \sqrt{t-4M_\pi^2} + (t - 2M_\pi^2 + 2\Delta^2) \arctan{\sqrt{t-4M_\pi^2}
\over 2 \Delta} \biggr] \theta(t-4M_\pi^2) \cr  & + {2C^2\over 3\pi}  \biggl[ -
\Delta \sqrt{t-4M_K^2} + (t - 2M_K^2 + 2\Delta^2) \arctan{\sqrt{t-4M_K^2}
\over 2 \Delta} \biggr] \theta(t-4M_K^2)\biggr\} \cr} \eqno(A.1)$$

$$\eqalign{ {\rm Im}&\sigma_{N}^{(j)}(t) = {M_K^2 \over 128  F_p^2 \sqrt
t} \biggl\{ 3  (D+F)^2 ({t\over 2} -M_\pi^2 ) \theta(t-4M_\pi^2) +
5({D\over 3}-F)^2({t\over 2} -M_\eta^2) \theta(t-4M_\eta^2)
\cr & + \xi_{\pi \eta}^{(j)}\,(D+F)(F-{D\over 3})(t-M_\pi^2
-M_\eta^2)\theta(t-(M_\pi+M_\eta)^2) +  \xi_K^{(j)}
(t-2M_K^2) \theta(t-4M_K^2)  \cr &  + {8 C^2 \over 3\pi } \biggl[ -
\Delta \sqrt{t-4M_\pi^2} + (t - 2M_\pi^2 + 2\Delta^2) \arctan{\sqrt{t-4M_\pi^2}
\over 2 \Delta} \biggr] \theta(t-4M_\pi^2) \cr  & + (2+\delta_{2j})
{4C^2\over 9\pi}  \biggl[ -
\Delta \sqrt{t-4M_K^2} + (t - 2M_K^2 + 2\Delta^2) \arctan{\sqrt{t-4M_K^2}
\over 2 \Delta} \biggr] \theta(t-4M_K^2) \biggr\}\cr} \eqno(A.2)$$
The shifts of the $\sigma$-terms from $t=0$ to the respective Cheng-Dashen
points
are most economically represented in the form of a once-subtracted
dispersion relation.
$$\eqalign{ \sigma_{\pi N}(2M_\pi^2)- \sigma_{\pi N}(0) &= {2M_\pi^2 \over \pi}
\int_{4M_\pi^2}^\infty dt {{\rm Im} \sigma_{\pi N}(t) \over t(t-2M_\pi^2)} \cr
 {\rm Re} \biggl[
 \sigma_{K N}^{(j)}(2M_K^2) - \sigma_{K N}^{(j)}(0) \biggr]
 & = {2M_K^2 \over
 \pi} \, \, {\cal P} \int_{4M_\pi^2}^\infty dt {{\rm Im}
 \sigma_{KN}^{(j)}(t) \over t(t-2M_K^2)} \cr }\eqno(A.3)$$
\bigskip \noindent
{\bf
REFERENCES}
\bigskip
\item{1.}J. Gasser and H. Leutwyler, {\it Ann. Phys. (N.Y.)\/}
 {\bf 158} (1984) 142;
 {\it Nucl. Phys.\/}
 {\bf B250} (1985) 465.
\smallskip
\item{2.}J. Gasser, M.E. Sainio and A. ${\check {\rm S}}$varc,
{\it Nucl. Phys.\/}
{\bf B307} (1988) 779.
\smallskip
\item{3.}E. Jenkins and A.V. Manohar, {\it Phys. Lett.\/} {\bf B255} (1991)
558.
\smallskip
\item{4.}J. Gasser and H. Leutwyler, {\it Phys. Reports\/} {\bf C87} (1982)
77.
\smallskip
\item{5.}S. Weinberg, {\it Nucl. Phys.\/} {\bf
B363} (1991) 3.
\smallskip
\item{6.}V. Bernard, N. Kaiser, J. Kambor
and Ulf-G. Mei{\ss}ner, {\it Nucl. Phys.\/} {\bf B388} (1992) 315.
\smallskip
\item{7.}E. Jenkins and A.V. Manohar, {\it Phys. Lett.\/} {\bf B259} (1991)
353.
\smallskip
\item{8.}E. Jenkins, {\it Nucl. Phys.\/}
{\bf B368} (1992) 190.
\smallskip
\item{9.}E. Jenkins, {\it Nucl. Phys.\/}
{\bf B375} (1992) 561.
\smallskip
\item{10.}M.N. Butler and M.J. Savage, {\it Phys. Lett.} {\bf B294} (1992)
369.
\smallskip
\item{11.}J. Bijnens, H. Sonoda and M.B. Wise, {\it Nucl. Phys.\/}
{\bf B261} (1985) 185.
\smallskip
\item{12.}J. Gasser, {\it Ann. Phys.\/} (N.Y.) {\bf
136} (1981) 62.
\smallskip
\item{13.}J. Gasser, H. Leutwyler and M.E. Sainio, {\it Phys. Lett.\/}
 {\bf 253B} (1991) 252, 260.
\smallskip
\item{14.}J.F. Donoghue and C.R. Nappi, {\it Phys. Lett.} {\bf B168} (1986)
105.
\smallskip
\item{15.}V. Bernard, R.L. Jaffe and Ulf--G. Mei{\ss}ner,  {\it Nucl.
Phys.\/} {\bf B308} (1988) 753.
\smallskip
\item{16.}E. Jenkins and A.V. Manohar, {\it Phys. Lett.\/} {\bf B281} (1992)
336.
\smallskip
\item{17.}
Ulf-G. Mei{\ss}ner,
{\it Int. J. Mod. Phys.}
{\bf E1} (1992) 561.
\smallskip
\item{18.}E. Jenkins and A.V. Manohar, in "Effective field theories of the
standard model", ed. Ulf--G. Mei{\ss}ner, World Scientific, Singapore,
1992.
\smallskip
\item{19.}
Ulf-G. Mei{\ss}ner, "Recent Developments in Chiral Perturbation Theory", Bern
University preprint BUTP-93/01, 1993.
\smallskip
\item{20.}H. Pagels and W. Pardee,
{\it Phys. Rev.\/} {\bf D4} (1971) 3225.
\smallskip
\item{21.}J. Gasser and A. Zepeda, {\it Nucl. Phys.\/}
{\bf B174} (1980) 445.
\smallskip
\item{22.}M. Bourquin et al., {\it Z. Phys.\/}
{\bf C21} (1983) 27.
\smallskip
\item{23.}R.L. Jaffe and A.V. Manohar, {\it Nucl. Phys.\/}
{\bf B337} (1990) 509.
\smallskip
\item{24.}R.L. Jaffe and C. Korpa,
{\it Comm. Nucl. Part.
Phys.\/} {\bf 17} (1987) 163.
\smallskip
\item{25.}I. Jameson, A.W. Thomas and G. Chanfray,
{\it J. Phys. G: Nucl. Part.
Phys.\/} {\bf 18} (1992) L159.
\smallskip
\item{26.}C. Ordonez and U. van Kolck, {\it Phys. Lett.\/} {\bf
B291} (1992) 459.
\smallskip
\item{27.}Ulf-G. Mei{\ss}ner,
{\it Comm. Nucl. Part.
Phys.\/} {\bf 20}
(1991) 119.
\smallskip
\item{28.}P.M. Gensini,
{\it J. Phys. G: Nucl. Part.
Phys.\/} {\bf 7} (1981) 1177.
\smallskip
\item{29.}P.M. Gensini, in $\pi N$ Newsletter no. 6, eds. R.E. Cutkowsky,
G. H\"ohler, W. Kluge and B.M.K. Nefkens, April 1992.
\bigskip
\goodbreak
\vfil
\eject

$$\hbox{\vbox{\offinterlineskip
\def\strut{\hbox{\vrule height 12pt depth 12pt width 0pt}}
\hrule
\halign{
\strut\vrule# \tabskip 0.0in &
\hfil#\hfil &
\hfil#\hfil &
\hfil#\hfil &
\vrule# &
\hfil#\hfil &
\hfil#\hfil &
\hfil#\hfil &
\hfil#\hfil &
\hfil#\hfil &
\hfil#\hfil &
\hfil#\hfil &
\hfil#\hfil &
\vrule# \tabskip 0.0in
\cr
&  $D$ &  $F$ &  $F_p$ && $b_D$ & $b_F$ & $b_0$ & $m_0$ &
$\sigma_{KN}^{(1)}(0)$ & $\sigma_{KN}^{(2)}(0)$ & SME & GMO
& \cr
& & & [MeV] && [GeV$^{-1}$] & [GeV$^{-1}$] & [GeV$^{-1}$]
& [GeV] & [MeV]
& [MeV] & [MeV] & [MeV] & \cr
\noalign{\hrule}
& 0.75 & 0.50 & 100 && 0.016 & -0.553 & -0.750 & 0.965 & 195.3 & 143.9
& -206 & 3.8 & \cr
& 0.756 & 0.477 & 100 && 0.037 & -0.540 & -0.753 & 0.958 & 204.3 & 146.0
& -192 & 2.3 & \cr
& 0.81 & 0.47 & 100 && 0.065 & -0.558 & -0.789 & 0.981 & 189.0 & 129.8
& -222 & 0.1 & \cr
& 0.75 & 0.50 & 92.6 && 0.008 & -0.610 & -0.788 & 1.014 & 154.0 & 117.4 &
-278 & 4.5 & \cr
& 0.75 & 0.50 & 112 && 0.027 & -0.483 & -0.703 & 0.904 & 245.8 & 176.4 &
-117 & 3.0 & \cr
\noalign{\hrule}}}}$$
\smallskip
{\noindent\narrower Table 1:\quad Results of the complete ${\cal O}(q^3)$
calculation. The values of $D,F$ and $F_p$ are input. GMO denotes the
combination $(3m_\Lambda + m_\Sigma - 2 m_N - 2 m_\Xi)/4$ of the octet baryon
masses. SME stands for the matrix element $m_s <p|\bar s s|p>$.
\smallskip}
\goodbreak
\bigskip

$$\hbox{\vbox{\offinterlineskip
\def\strut{\hbox{\vrule height 12pt depth 12pt width 0pt}}
\hrule
\halign{
\strut\vrule# \tabskip 0.0in &
\hfil#\hfil &
\hfil#\hfil &
\hfil#\hfil &
\hfil#\hfil &
\vrule# &
\hfil#\hfil &
\hfil#\hfil &
\hfil#\hfil &
\hfil#\hfil &
\hfil#\hfil &
\hfil#\hfil &
\hfil#\hfil &
\hfil#\hfil &
\vrule# \tabskip 0.0in
\cr
&  $D$ &  $F$ &  $\Delta$ & $C$ && $b_D$ & $b_F$ & $b_0$ & $m_0$ &
$\sigma_{KN}^{(1)}(0)$ & $\sigma_{KN}^{(2)}(0)$ & SME & GMO
& \cr
& & & [MeV] & && [GeV$^{-1}$] & [GeV$^{-1}$] & [GeV$^{-1}$]
& [GeV] & [MeV]
& [MeV] & [MeV] & [MeV] & \cr
\noalign{\hrule}
& 0.75 & 0.50 & 293 &1.8 && 0.623 & -1.057 & -1.744 & 1.330 & -39.4 & -111.5
& -667 & 11.5 & \cr
& 0.75 & 0.50 & 293 & 1.5 && 0.438 & -0.903 & -1.441 & 1.218 & 32.3 & -33.4
& -526 & 9.2 & \cr
& 0.75 & 0.50 & 231 & 1.8 && 0.642 & -1.072 & -1.752 & 1.360 & -58.2 & -131.6
& -704 & 12.0 & \cr
& 0.75 & 0.50 & 231 & 1.5 && 0.451 & -0.914 & -1.446 & 1.239 & 19.3 & -47.4 &
-551 & 9.5 & \cr
& 0.56 & $2D/3$ & 293 & $2D$ && 0.273 & -0.596 & -1.033 & 0.975 & 214.3 & 115.6
& -192 & 5.1 & \cr
& 0.56 & $2D/3$ & 231 & $2D$ && 0.281 & -0.602 & -1.036 & 0.986 & 207.0 & 107.8
& -206 & 5.3 & \cr
\noalign{\hrule}}}}$$
\smallskip
{\noindent\narrower Table 2:\quad Results of the calculation including the full
decuplet intermediate states. The values of $D,F,\Delta$ and $C$ are input.
\smallskip}
\goodbreak
\bigskip

$$\hbox{\vbox{\offinterlineskip
\def\strut{\hbox{\vrule height 12pt depth 12pt width 0pt}}
\hrule
\halign{
\strut\vrule# \tabskip 0.0in &
\hfil#\hfil &
\hfil#\hfil &
\hfil#\hfil &
\hfil#\hfil &
\vrule# &
\hfil#\hfil &
\hfil#\hfil &
\hfil#\hfil &
\hfil#\hfil &
\hfil#\hfil &
\hfil#\hfil &
\hfil#\hfil &
\hfil#\hfil &
\vrule# \tabskip 0.0in
\cr
&  $D$ &  $F$ &  $\Delta$ & $C$ && $b_D$ & $b_F$ & $b_0$ & $m_0$ &
$\sigma_{KN}^{(1)}(0)$ & $\sigma_{KN}^{(2)}(0)$ & SME & GMO
& \cr
& & & [MeV] & && [GeV$^{-1}$] & [GeV$^{-1}$] & [GeV$^{-1}$]
& [GeV] & [MeV]
& [MeV] & [MeV] & [MeV] & \cr
\noalign{\hrule}
& 0.75 & 0.50 & 293 &1.8 && 0.547 & -0.995 & -1.696 & 1.222 & 38.2 & -23.3
& -513 & 7.0 & \cr
& 0.75 & 0.50 & 293 & 1.5 && 0.385 & -0.860 & -1.407 & 1.143 & 86.2 & 27.8
& -419 & 6.0 & \cr
& 0.75 & 0.50 & 231 & 1.8 && 0.518 & -0.971 & -1.673 & 1.185 & 67.0 & 10.3
& -455 & 4.8 & \cr
& 0.75 & 0.50 & 231 & 1.5 && 0.365 & -0.843 & -1.391 & 1.117 & 106.1 & 51.1
& -379 & 4.5 & \cr
& 0.56 & $2D/3$ & 293 & $2D$ && 0.244 & -0.572 & -1.014 & 0.933 & 244.3 & 149.7
& -132 & 3.4 & \cr
& 0.56 & $2D/3$ & 231 & $2D$ && 0.233 & -0.563 & -1.005 & 0.919 & 255.4 & 162.8
& -110 & 2.5 & \cr
\noalign{\hrule}}}}$$
\smallskip
{\noindent\narrower Table 3:\quad  Same as in table 2 but with the decuplet
contributions expanded up to and including ${\cal O}(q^4)$.
\smallskip}
\goodbreak
\bigskip
\end